\documentclass[pra,aps,preprint,showpacs]{revtex4}
\newcommand{\ket}[1]{|#1\rangle}
\newcommand{\bra}[1]{\langle #1|}

\begin{document}
\title{ Quantitative conditions do not guarantee the validity of the
    adiabatic approximation}
\author{D. M. Tong$^{1,2}$
\footnote{Electronic address: phytdm@nus.edu.sg} , K. Singh$^1$,
L. C. Kwek$^{1,3}$,  and C. H. Oh$^1$ \footnote{Electronic
address: phyohch@nus.edu.sg} } \affiliation{ $^1$ Department of
Physics, National University of
Singapore, 10 Kent Ridge Crescent, Singapore 119260, Singapore \\
$^2$Department of Physics, Shandong Normal University, Jinan 250014, P. R. China\\
$^3$National Institute of Education, Nanyang Technological
University, 1 Nanyang Walk, Singapore 639798, Singapore
}
\date{\today}
\begin{abstract}
In this letter, we point out that the widely used quantitative conditions in the adiabatic theorem are insufficient in that they do not guarantee the validity of the adiabatic approximation. We also reexamine the inconsistency issue raised by Marzlin and Sanders (Phys. Rev. Lett. 93, 160408, 2004) and elucidate the underlying cause.
\end{abstract}
\maketitle
\date{\today}

The adiabatic theorem\cite{Ehrenfest,Born,Schwinger,Kato} is one of the basic results in quantum theory. It has been widely applied in both theories and experiments\cite{Landau,Zener,Gell,Berry,Farhi} and has grown in importance in the recent years due to a number of extensions and applications\cite{Avron,Ali,Erik,Lidar,Roland,Andris,Sun,Tong}. The validity of the application of the theorem had never been doubted until Marzlin and Sanders recently claimed that the application of adiabatic theorem may lead to an inconsistency\cite{Marzlin}. Essentially they demonstrated that the application of the adiabatic theorem implied a non-unit norm in the states. There have been some attempts at addressing this problem. For instance some authors\cite{Pati} proposed revolving the inconsistency by dropping some nondiagonal terms in the transition amplitudes; but in doing so, ended up with another inconsistency. Yet, others\cite{Sarandy,Wu,Tong6} have suggested that the inconsistency does not arise from the adiabatic theorem itself, but is a result of incorrect manipulations in mathematics.
However, we now find that the inconsistency may exist in the use of the adiabatic approximation. It is actually a reflection of a more crucial issue that the quantitative adiabatic conditions are insufficient. In this letter, we show that the widely used quantitative statements of the adiabatic conditions in the adiabatic theorem, which are often deemed as sufficient, are really insufficient. They cannot sufficiently guarantee the validity of the adiabatic approximation.

Before proceeding further, it is instructive to recapitulate the statement of the adiabatic theorem. The theorem states that if a quantum system with a time-dependent nondegenerate Hamiltonian $H(t)$ is initially in $n$-th eigenstate of $H(0)$, and if $H(t)$ evolves slowly enough, then the state of the system at time $t$ will remain in the $n$-th instantaneous eigenstate of  $H(t)$ up to a multiplicative phase factor. In the literature, the term ``$H(t)$ evolves slowly enough", is usually encoded in the quantitative requirement that\cite{Tong3}
\begin{eqnarray}
\frac{\left |\bra{E_m(t)}\dot H(t)\ket{E_n(t)}\right |}{\left |E_m(t)-E_n(t)\right |^2}\ll 1,~m\neq n,~~t\in[0,T],
\label{cons0}
\end{eqnarray}
or equivalently
\begin{eqnarray}
\left |\frac{\langle{E_m(t)}\ket{\dot E_n(t)}}{E_m(t)-E_n(t)}\right | \ll 1,~m\neq n,~~t\in[0,T],
\label{cons}
\end{eqnarray}
where $E_m(t)$ and $\ket{E_m(t)}$ are the entirely discrete and nondegenerate instantaneous eigenvalues and eigenstates of $H(t)$, and $T$ is the total evolution time.

We will show that the quantitative adiabatic conditions expressed
by Eq. (\ref{cons}) is not sufficient in guaranteeing the validity
of the adiabatic approximation. To this end, we consider two
related $N$-dimensional quantum systems $S^a$ and $S^b$, which are
defined by the Hamiltonians $H^a(t)$ and $H^b(t)$ respectively.
The two systems are related through
\begin{eqnarray}
H^b(t)=-U^{a\dag}(t)H^a(t)U^a(t),
 \label{hbaa}
\end{eqnarray}
which means that the evolution operator for $S^b$ is the Hermitian
conjugate of the evolution operator for $S^a$.

We first consider the system $S^a$. The instantaneous eigenvalues
and normalized eigenstates of $H^a$ are denoted as $E^a_m(t)$ and
$\ket{E^a_m(t)}$ respectively, which satisfy
\begin{eqnarray}
H^a(t)\ket{E^a_m(t)}=E^a_m(t)\ket{E^a_m(t)},~m=1,\ldots,N.
\label{ha}
\end{eqnarray}
We assume that $E^a_m(t)$ are entirely discrete and nondegenerate and they fulfill the adiabatic conditions
\begin{eqnarray}
\left |{\frac{\langle E^a_m(t)\ket{\dot E^a_n(t)}}{E^a_m(t)-E^a_n(t)}}\right |\ll 1,
~m\neq n,~~t\in[0,T].
\label{consa}
\end{eqnarray}
Now, the state of the system, denoted by $\ket{\psi^a(t)}$, is the solution of the Schr\"odinger equation
\begin{eqnarray}
i\frac{d}{dt}\ket{\psi^a(t)}=H^a(t)\ket{\psi^a (t)},
\label{schra}
\end{eqnarray}
with the initial state $\ket{\psi^a(0)}$. If the initial state is in the $n$-th eigenstate $\ket{E^a_n(0)}$, then the state at time $t$, $\ket{\psi^a(t)}$, can be expressed exactly as
\begin{eqnarray}
\ket{\psi^a(t)}&=U^a(t)\ket{E^a_n(0)},
\label{psian}
\end{eqnarray}
where the unitary operator $U^a(t)=\text{T}exp\left(-i\int_0^tH^a(t')dt'\right)$.
In the application of the adiabatic theorem, we have
\begin{eqnarray}
\ket{\psi^a(t)}\approx\ket{\psi^a_{adi}(t)},
\label{psita}
\end{eqnarray}
where the state $\ket{\psi^a_{adi}(t)}$ is given as
\begin{eqnarray}
\ket{\psi^a_{adi}(t)}=e^{i\alpha^a_n (t)}\ket{E^a_n(t)}
\label{psita2}
\end{eqnarray}
with the phase $\alpha^a_n(t)$ taking the form
\begin{eqnarray}
\alpha^a_n(t)=-\int_0^tE^a_n(t')dt'+i\int_0^t\langle{E^a_n(t')}\ket{\dot E^a_n(t')}dt'.
\label{alphat}
\end{eqnarray}
Noticing that $|\bra{\psi^a_{adi}(t)}\psi^a(t)\rangle |=| \bra{E^a_n(t)}U^a(t)\ket{E^a_n(0)}|$, one concludes that the approximation (\ref{psita}) is acceptable if and only if
\begin{eqnarray}
|\bra{E^a_n(t)}U^a(t)\ket{E^a_n(0)}|\approx 1.
\label{bbb}
\end{eqnarray}

Next we consider the second quantum system $S^b$, defined by the Hamiltonian $H^b(t)$. We enumerate its instantaneous eigenvalues $E^b_m(t)$ and normalized eigenstates $\ket{E^b_m(t)}$ through
\begin{eqnarray}
H^b(t)\ket{E^b_m(t)}&=&E^b_m(t)\ket{E^b_m(t)},~m=1,\ldots,N,
\label{hb}
\end{eqnarray}
For system $S^b$, the state $\ket{\psi^b(t)}$ is governed by the Schr\"odinger equation
\begin{eqnarray}
i\frac{d}{dt}\ket{\psi^b(t)}=H^b(t)\ket{\psi^b (t)},
\label{schra}
\end{eqnarray}
with the initial state $\ket{\psi^b(0)}$. If the system is initially in the $n$-th eigenstate $\ket{E^b_n(0)}$, $\ket{\psi^b(t)}$ can be expressed exactly as
\begin{eqnarray}
\ket{\psi^b(t)}={U^b}(t)\ket{E^b_n(0)},
\label{psibn}
\end{eqnarray}
where $U^b(t)=\text{T}exp\left(-i\int_0^tH^b(t')dt'\right)$.

Under relation (\ref{hbaa}), it is easy to see that there is a one-to-one corresponding between the eigenvalues and eigenstates of the two systems:
\begin{eqnarray}
&E^b_n(t)&=-E^a_n(t),
\label{ebnt}\\
&\ket{E^b_n(t)}&=U^{a\dag}(t)\ket{E^a_n(t)}.
\label{kebnt}
\end{eqnarray}
From this correspondence, we note that
\begin{widetext}
\begin{eqnarray}
\langle E^b_m(t)\ket{\dot E^b_n(t)}&=&\bra{E^a_m(t)}U^a(t){\dot U}^{a\dag}(t)\ket{E^a_n(t)} +
\bra{ E^a_m(t)}U^a(t){U^a}^\dag(t)\ket{\dot E^a_n(t)} \nonumber\\
&=& i\bra{ E^a_m(t)}H^a(t) \ket{E^a_n(t)} +\bra{ E^a_m(t)}\dot E^a_n(t)\rangle \nonumber\\
&=& iE^a_m(t)\delta_{mn} +\bra{ E^a_m(t)}\dot E^a_n(t)\rangle,
\label{EE}
\end{eqnarray}
\end{widetext}
which implies
\begin{eqnarray}
\left |\frac{\langle E^b_m(t)\ket{\dot E^b_n(t)}}{ E^b_m(t)-E^b_n(t)}\right |
=\left |{\frac{\langle E^a_m(t)\ket{\dot E^a_n(t)}}{E^a_m(t)-E^a_n(t)}}\right |\ll 1,~~m\neq n,~~t\in[0,T].
\label{cons2}
\end{eqnarray}
Eq. (\ref{cons2}) shows that the system $S^b$ satisfies the adiabatic conditions  if and only if  $S^a$ satisfies them. We may now apply the adiabatic theorem to $\ket{\psi^b(t)}$, so that
\begin{eqnarray}
\ket{\psi^b(t)}\approx\ket{\psi^b_{adi}(t)},
\label{psib2}
\end{eqnarray}
where the state $\ket{\psi^b_{adi} (t)}$ is given as
\begin{eqnarray}
\ket{\psi^b_{adi} (t)}= e^{i\alpha_n^b(t)}\ket{E^b_n(t)},
\label{psiba}
\end{eqnarray}
with $\alpha_n^b(t)$ taking the form
\begin{eqnarray}
\alpha_n^b(t)&=& -\int_0^t E^b_n(t')dt'+i\int_0^t\langle E^b_n(t')\ket{\dot E^b_n(t')}dt'\nonumber\\
&=&i\int_0^t\bra{E^a_n(t')}\dot E^a_n(t')\rangle dt'.
\label{alphat2}
\end{eqnarray}
We now calculate the fidelity $|\bra{\psi^b_{adi}(t)}\psi^b(t)\rangle|$. From Eqs. (\ref{hbaa}) and (\ref{kebnt}), we obtain the relations $U^b(t)=U^{a\dag}(t)$ and $\ket{E^b_n(0)}=\ket{E^a_n(0)}$, and in using these relations we get
\begin{eqnarray}
\left |\bra{\psi^b_{adi}(t)}\psi^b(t)\rangle \right|=\left | \bra{E^a_n(t)}E^a_n(0)\rangle \right |.
\label{nin2}
\end{eqnarray}
Eq. (\ref{nin2}) shows that the approximation (\ref{psib2}) is acceptable if and only if
\begin{eqnarray}
|\bra{E^a_n(t)}E^a_n(0)\rangle|\approx 1.
\label{aaa}
\end{eqnarray}

Comparing Eqs. (\ref{bbb}) and (\ref{aaa}), one finds that the two expressions are quite different in general. They may not hold at the same time except for some special cases, which means that the two approximate equations (\ref{psita}) and (\ref{psib2}) may not be always valid. Thus, for a given quantum system $S^a$, one can always construct another quantum system $S^b$, with both of them fulfilling the same adiabatic conditions. The fact that $E^i_n(t)$ and $\ket{E^i_n(t)}~(i=a,b)$ do satisfy the conditions (\ref{cons2}) but $\ket{\psi^a(t)}$ or $\ket{\psi^b(t)}$ may not approximate to $\ket{\psi^a_{adi}(t)}$ or $\ket{\psi^b_{adi}(t)}$ indicates that the adiabatic conditions (\ref{cons}) do not sufficiently guarantee the validity of the adiabatic approximation. Our analysis clearly suggests that the adiabatic conditions described by Eq. (\ref{cons}) is insufficient.

Further to substantiate the result obtained above, we now furnish an example to illustrate the fact that the approximation $\ket{\psi^b(t)}\approx\ket{\psi^b_{adi}(t)}$ is invalid while  $\ket{\psi^a(t)}\approx\ket{\psi^a_{adi}(t)}$ is valid. To this end, consider the well-known model, a spin-half particle in a rotating magnetic field. We denote the system as $S^a$. The Hamiltonian of the system is
\begin{eqnarray}
H^a(t)&=&-\frac{\omega_0}{2}(\sigma_x\sin\theta\cos\omega t+\sigma_y\sin\theta\sin\omega t+\sigma_z\cos\theta)
\end{eqnarray}
where $\omega_0$ is a time-independent parameter defined by the magnetic moment of the spin and the intensity of external magnetic field, $\omega $ is the rotating frequency of the magnetic field and $\sigma_i,~i=x,y,z,$ are Pauli matrices. The instantaneous eigenvalues and eigenstates of $H^a(t)$ are
\begin{eqnarray}
E^a_1(t)=\frac{\omega_0}{2}, ~~~~\ket{E^a_1(t)}=\left(\begin{array}{c}
e^{-i\omega t/2}\sin\frac{\theta}{2}\\-e^{i\omega t/2}\cos\frac{\theta}{2}
\end{array}\right);
\end{eqnarray}
\begin{eqnarray}
E^a_2(t)=-\frac{\omega_0}{2},~~~~\ket{E^a_2(t)}=\left(\begin{array}{c}
e^{-i\omega t/2}\cos\frac{\theta}{2}\\e^{i\omega t/2}\sin\frac{\theta}{2}
\end{array}\right).
\label{ketE2}
\end{eqnarray}
From $H^a(t)$, we may construct another quantum system $S^b$ defined by the Hamiltonian $H^b(t)=-U^{a\dag}(t)H^a(t)U^a(t)$. The eigenvalues and eigenstates of  $H^b(t)$ are
\begin{eqnarray}
E^b_1(t)=-\frac{\omega_0}{2}, ~~~~\ket{E^b_1(t)}=U^{a\dag}(t)\ket{E^a_1(t)};
\label{ketE3}
\end{eqnarray}
\begin{eqnarray}
E^b_2(t)=\frac{\omega_0}{2},
~~~~\ket{E^b_2(t)}=U^{a\dag}(t)\ket{E^a_2(t)}.
\label{ketE4}
\end{eqnarray}
It is easy to show that the adiabatic conditions (\ref{cons2}) are satisfied as long as $\omega_0\gg \omega\sin\theta$.

Suppose that the system $S^b$ is initially in the state $\ket{E^b_1(0)}$. we now calculate   $\ket{\psi^b(t)}$ defined by Eq. (\ref{psibn}) and $\ket{\psi^b_{adi}(t)}$ defined by Eq. (\ref{psib2}). To this end, we first need to evaluate the unitary operator $U^a(t)$, and we obtain
\begin{widetext}
\begin{eqnarray}
U^a(t)=\left(\begin{array}{cc}
(\cos\frac{\overline\omega t}{2}+i\frac{\omega+\omega_0\cos\theta}{\overline\omega}\sin\frac{\overline\omega t}{2})e^{-i\frac{\omega t}{2}} & i\frac{\omega_0\sin\theta}{\overline\omega}\sin\frac{\overline\omega t}{2}e^{-i\frac{\omega t}{2}}\\
i\frac{\omega_0\sin\theta}{\overline\omega}\sin\frac{\overline\omega t}{2}e^{i\frac{\omega t}{2}}
&(\cos\frac{\overline\omega t}{2}-i\frac{\omega+\omega_0\cos\theta}{\overline\omega}\sin\frac{\overline\omega t}{2})e^{i\frac{\omega t}{2}} \end{array}\right),
\label{u}
\end{eqnarray}
\end{widetext}
where $\overline\omega=\sqrt{\omega_0^2+\omega^2+2\omega_0\omega
\cos\theta}$. With $U^a(t)$ found, we obtain the exact state $\ket{\psi^b(t)}$ of the state  and the approximate state $\ket{\psi^b_{adi}(t)}$  as implied by the adiabatic theorem:
\begin{widetext}
\begin{eqnarray}
\ket{\psi^b(t)}
 = \left(\begin{array}{c}(\cos\frac{\overline\omega
t}{2}-i\frac{\omega+\omega_0\cos\theta}{\overline\omega}\sin\frac{\overline\omega
t}{2})\sin\frac{\theta}{2} e^{i\frac{\omega t}{2}}+
i\frac{\omega_0\sin\theta}{\overline\omega}\sin\frac{\overline\omega
t}{2}\cos\frac{\theta}{2} e^{-i\frac{\omega t}{2}} \\
-(\cos\frac{\overline\omega
t}{2}+i\frac{\omega+\omega_0\cos\theta}{\overline\omega}\sin\frac{\overline\omega
t}{2})\cos\frac{\theta}{2} e^{-i\frac{\omega t}{2}}-
i\frac{\omega_0\sin\theta}{\overline\omega}\sin\frac{\overline\omega
t}{2}\sin\frac{\theta}{2}e^{i\frac{\omega t}{2}}
\end{array}\right),
\label{e1}
\end{eqnarray}
\begin{eqnarray}
\ket{\psi^b_{adi}(t)}
=e^{-i\frac{\omega\cos\theta t}{2}}\left(\begin{array}{c}(\cos\frac{\overline\omega
t}{2}-i\frac{\omega+\omega_0\cos\theta}{\overline\omega}\sin\frac{\overline\omega
t}{2})\sin\frac{\theta}{2} +
i\frac{\omega_0\sin\theta}{\overline\omega}\sin\frac{\overline\omega
t}{2}\cos\frac{\theta}{2}  \\
-(\cos\frac{\overline\omega
t}{2}+i\frac{\omega+\omega_0\cos\theta}{\overline\omega}\sin\frac{\overline\omega
t}{2})\cos\frac{\theta}{2} -
i\frac{\omega_0\sin\theta}{\overline\omega}\sin\frac{\overline\omega
t}{2}\sin\frac{\theta}{2}
\end{array}\right).
\label{e2}
\end{eqnarray}
\end{widetext}
We see that $\ket{\psi^b(t)}$ is quite different from $\ket{\psi^b_{adi}(t)}$. The fidelity between them is
\begin{eqnarray}
\left |\bra{\psi^b_{adi}(t)}\psi^b (t)\rangle \right |^2=1- \sin^2\theta\sin^2\frac{\omega t}{2},
\label{e3}
\end{eqnarray}
which shows that $\ket{\psi^b(t)}$ cannot approximate to $\ket{\psi^b_{adi}(t)}$ although  $\omega_0\gg\omega\sin\theta$ is fulfilled. However, for this model the approximation (\ref{psita}) is valid\cite{Tong}. This example illustrates the fact that fulfillment of the adiabatic conditions (\ref{cons}) does not sufficiently guarantee the validity of the adiabatic approximation. Another solvable example can be found in \cite{Marzlin}.

It is worth noting that there are alternative expressions for the adiabatic conditions in the literature. One may wonder whether other quantitative expressions of the conditions are sufficient. Here, we will examine two commonly used ones.

One version\cite{Lidar} expresses the adiabatic conditions as
\begin{eqnarray}
\max\left |\frac{\bra{E_m(t)}\dot H(t) \ket{E_n}(t)}{E_n(t)-E_m(t)}\right | \ll \min\left |E_n(t)-E_m(t)\right |,~~m\neq n,~~t\in[0,T],
\label{cons7}
\end{eqnarray}
where $\max|f(t)|$ ($\min |f(t)|$) means the maximum (minimum) value of $|f(t)|, ~t\in[0,T]$. In the use of Eqs. (\ref{ebnt}) and (\ref{EE}), one may immediately infer that Eq. (\ref{cons7}) is satisfied by $S^b$ if and only if it is satisfied by $S^a$. This in turn will lead to the same result as expressed in Eqs. (\ref{bbb}) and  (\ref{aaa}), which means that the present conditions are insufficient too.

Another commonly used version\cite{Roland} of the adiabatic theorem states if
\begin{eqnarray}
\frac{\max\left |\bra{E_m(t)}\dot H(t) \ket{E_n(t)}\right |}{\min\left |E_m(t)-E_n(t)\right |^2} \leq \varepsilon,~m\neq n,~~t\in[0,T],
\label{cons8}
\end{eqnarray}
then
\begin{eqnarray}
\left |\bra{\psi^{adi}(T)}\psi(T)\rangle\right |\geq 1-\varepsilon^2.
\label{resu}
\end{eqnarray}
Similarly, we note that the conditions (\ref{cons8}) cannot guarantee the validity of Eq. (\ref{resu}). In fact, using Eqs. (\ref{ebnt}) and (\ref{EE}), one finds that Eq. (\ref{cons8}) is satisfied by $S^b$ if and only if it is satisfied by $S^a$. We then again arrive at Eqs. (\ref{bbb}) and (\ref{aaa}). To further elaborate on this, we can use the previous example to illustrate the point. For the Hamiltonian $H^b(t)$ in that example, the left side of Eq. (\ref{cons8}) is equal to $\omega\sin\theta /\omega_0$, which may be very small, but the left side of Eq. (\ref{resu}) is $[1-\sin^2(\theta/2)\sin^2(\omega T/2)]$, which may not approximate to $1$ in general.

It is instructive to reexamine the inconsistency raised by Marzlin and Sanders in light of our work. In the following, we give an alternative proof to the inconsistency, which can avoid the criticism levelled against their proof. Indeed, by using Eqs. (\ref{psib2}) and (\ref{psiba}), we may immediately get
\begin{widetext}
\begin{eqnarray}
\bra{E^a_n(0)}U(t)U^\dag \ket{E^a_n(0)}&=&\bra{E^a_n(0)}U(t)\ket{\psi^b(t)}\nonumber\\
&\approx&\bra{E^a_n(0)}U(t)\ket{\psi^b_{adi}(t)}\nonumber\\
&=& e^{-\int_0^t\bra{E^a_n(t')}\dot E^a_n(t')\rangle dt'}\langle E^a_n(0)U(t)U^\dag(t)\ket{E^a_n(t)}\nonumber\\
&=& e^{-\int_0^t\bra{E^a_n(t')}\dot E^a_n(t')\rangle dt'}\langle E^a_n(0)\ket{E^a_n(t)}\nonumber\\
&\neq& 1.
\label{nin5}
\end{eqnarray}
\end{widetext}
This is the inconsistency raised in \cite{Marzlin}.
It may be worth noting that, we have only used the adiabatic approximation (\ref{psib2}) once and all other calculations are exact. So, we can say that the inconsistency claimed by Marzlin and Sanders does exist in the use of the adiabatic approximation. The essential reason for the inconsistency is the insufficient adiabatic conditions, which cannot guarantee the validity of the adiabatic approximation.

Before concluding, to give a simple physical picture may be helpful for comprehending the result that the adiabatic conditions are satisfied but the adiabatic approximation may be invalid. To this end, we consider a Hamiltonian that can be written as a sum of two parts, the base part and the perturbing part. In the case where the perturbing part is a periodic rapid varying perturbation in resonance with the base Hamiltonian, the effect of the perturbing part to the system may accumulate to an appropriate scale after a long time, no matter how small the perturbation is, and the transition may be driven between two eigenstates. In this case, the adiabatic approximation is clearly invalid, but the conditions (\ref{cons}) may be satisfied as long as the perturbation is small enough. So, for such a system, at least , the adiabatic conditions may fail to guarantee the validity of the adiabatic approximation.

In conclusion, we have shown that the widely used quantitative statements of the adiabatic conditions, such as (\ref{cons}), (\ref{cons7}) and (\ref{cons8}), are insufficient in guaranteeing the validity of the adiabatic approximation. This implies that fulfilling only the quantitative statements cannot meet the adiabatic criterion that is required by the adiabatic theorem. Besides, we have reinterpreted the inconsistency raised by Marzlin and Sanders  and found that the essential reason of leading to the inconsistency is  the use of the insufficient adiabatic conditions. In passing, we will like to add that the work presented here reopens the all-important question as to the right quantitative sufficiency conditions that will mirror an adiabatic evolution.

\vskip 0.3 cm Tong would like to thank Dr. E. Sj\"oqvist for his
valuable comments. The work was supported by NUS Research Grant
No. R-144-000-071-305.


\begin{thebibliography}{99}
\bibitem{Tong3}There are many other alternative expressions of the conditions. For simplicity, we choose the simple one without loss of generality.
\end{thebibliography}
\end{document}